\documentclass[journal]{IEEEtran}

\usepackage{graphicx}
\usepackage{subfig}
\usepackage{amssymb,amsmath,bm}
\usepackage{booktabs}

\newcommand{\bs}[1]{\boldsymbol {#1}}

\newcommand{\argmax}{\mathop{\rm argmax}\limits}

\ifCLASSINFOpdf
\else
\fi
\begin{document}
%
\title{Complex-Valued Restricted Boltzmann Machine for Direct Speech Parameterization from Complex Spectra}
%
%
%

\author{Toru~Nakashika,~\IEEEmembership{Member,~IEEE,}, Shinji~Takaki,~\IEEEmembership{Member,~IEEE}
        and~Junichi~Yamagishi,~\IEEEmembership{Senior Member,~IEEE}
\thanks{T. Nakashika is with the Graduate School
of Informatics and Engineering, the University of Electro-Communications, Tokyo,
Japan e-mail: nakashika@uec.ac.jp}
\thanks{S. Takaki and J. Yamagishi are with National Institute of Informatics,
Tokyo, Japan
e-mail: takaki@nii.ac.jp, jyamagis@nii.ac.jp}
\thanks{Manuscript received April 19, 2005; revised December 27, 2012.}}

%
%

\markboth{Journal of \LaTeX\ Class Files,~Vol.~11, No.~4, December~2012}%
{Shell \MakeLowercase{\textit{et al.}}: Bare Demo of IEEEtran.cls for Journals}
%



\maketitle

\begin{abstract}
This paper describes a novel energy-based probabilistic distribution that represents complex-valued data and explains how to apply it to direct feature extraction from complex-valued spectra.
The proposed model, the complex-valued restricted Boltzmann machine (CRBM), is designed to deal with complex-valued visible units as an extension of the well-known restricted Boltzmann machine (RBM).
Like the RBM, the CRBM learns the relationships between visible and hidden units without having connections between units in the same layer, which dramatically improves training efficiency by using Gibbs sampling or contrastive divergence (CD).
Another important characteristic is that the CRBM also has connections between real and imaginary parts of each of the complex-valued visible units that help represent the data distribution in the complex domain.
In speech signal processing, classification and generation features are often based on amplitude spectra (e.g., MFCC, cepstra, and mel-cepstra) even if they are calculated from complex spectra, and they ignore  phase information.
In contrast, the proposed feature extractor using the CRBM directly encodes the complex spectra (or another complex-valued representation of the complex spectra) into binary-valued latent features (hidden units).
Since the visible-hidden connections are undirected, we can also recover (decode) the complex spectra from the latent features directly.
Our speech coding experiments demonstrated that the CRBM outperformed other speech coding methods, such as methods using the conventional RBM, the mel-log spectrum approximate (MLSA) decoder, etc.
\end{abstract}


%
\IEEEpeerreviewmaketitle

\section{Introduction}
\label{sec:intro}
%
%
%
%
\IEEEPARstart{D}{eep learning} is one of the recent hottest topics in a wide range of research fields, such as artificial intelligence, machine learning, and signal processing that includes image classification, speech recognition, etc\cite{LeCun:2015dt}.
Many models have been proposed as tools of deep learning; one of the most widely-used and famous models is a deep belief-net (DBN) \cite{hinton2006fast} that stacks multiple restricted Boltzmann machines (RBMs) layer-by-layer.
The RBM is a probabilistic model that consists of visible and hidden units and has often been used alone as a feature extractor, a generator, and as a classifier and pre-training scheme of deep neural networks.
Many extensions of the RBM have been proposed for task specification \cite{Salakhutdinov:2009uo,Krizhevsky:2010va,Sohn:2013vc,Nakashika:2016gi}.
Although the RBM has been used in many tasks, the RBM traditionally identified visible units as either binary-valued or real-valued \cite{hinton2006fast,Freund:1994tu,Lee:2008uz}.

Representations based on the amplitude spectra of speech (such as MFCC, cepstra, and mel-cepstra) are traditionally used in speech signal processing as input features of speech recognition or output features of speech synthesis because the amplitude spectra are more effective and relevant to our auditory field for such tasks than phase spectra.
Raw amplitude spectral representation can be also used \cite{takaki2017direct,palaz2015analysis}.
However, these features that include the amplitude spectra theoretically lack phase information, and single use of the amplitude-based features cannot completely recover the original complex spectra with reasonable computational resources easily, even when using the well-known Griffin-Lim algorithm \cite{Griffin:1984bl}.
In \cite{oord2016wavenet,oord2017neural,kobayashi2014statistical}, it is reported that the generated speech signals from direct waveform modification or synthesis are much more natural than those from methods that are based on phase reconstruction from amplitude spectra.
Furthermore, there are many cases in other kinds of signal processing in which we have to deal with complex-valued actual data such as fMRI images, wireless signals, acoustic intensity, etc.
Other machine learning models---that is, neural networks, Boltzmann machines, and non-negative matrix factorization (NMF) \cite{lee1999learning}---have their extensions proposed to represent complex-valued data \cite{nemoto1992complex,Zemel:1995ga,Kameoka:2009kx}.

In our previous work \cite{nakashika2017complex}, we proposed an extended model of the RBM, namely ``complex-valued RBM (CRBM), '' to tackle representing complex-valued data in the RBM-based approach in particular.
The CRBM includes three important characteristics.
Firstly, the CRBM has no connections across dimensions in the same layers but has connections between visible and hidden units like the RBMs.
These restrictions make it exceedingly easy to estimate the parameters using Gibbs sampling or CD \cite{hinton2006fast}, which cannot be seen in an extension of a Boltzmann machine (directional-unit Boltzmann machine (DUBM) \cite{Zemel:1995ga}) that feeds complex-valued data that has connections across dimensions and has difficulties in parameter estimation.
Secondly, unlike the conventional RBM, the CRBM restricts the connections between different visible units but still has connections between real and imaginary parts of each visible unit.
Therefore, the CRBM represents the complex-valued data distribution more accurately than the RBMs, especially when there are correlations between the real and imaginary parts.
Thirdly, the CRBM represents the complex-valued visible units in a rectangular form that consists of real and imaginary components, while traditional representation methods of complex-valued data that include a DUBM are based on a polar form of phase and amplitude components.
We can generate samples from the distribution straightforwardly in the CRBM.
The conditional probability of visible units given hidden units form a complex-normal distribution, which makes the real and imaginary components Gaussian-distributed.
In \cite{nakashika2017complex}, we showed that the CRBM sufficiently recovered the amplitude and phase components as well as the real and imaginary components in the speech coding experiments.

We also propose some improvements and learning techniques for the CRBM-based speech parameterization.
First, we reduce the number of dimensions by feeding complex-valued visible features obtained by the complex principal component analysis (CPCA) \cite{horel1984complex} into the CRBM instead of the raw complex spectra.
Next, we employ the maximum likelihood parameter generation (MLPG) \cite{tokuda2000speech} to generate the trajectories of the CPCA features for better representation of speech sequences.
Finally, we extend the Adam algorithm to deal with the complex-valued parameters (referred to as ``complex Adam'' or ``CAdam''), which makes convergence of the model training faster than the steepest descent/ascent.
In the experiments, we compare the performance of the improved CRBM method with other speech coding methods, such as the conventional RBM, the mel-log spectrum approximate (MLSA), etc.

This paper is organized as follows.
In Section 2, we briefly review the conventional RBM and we present our proposed model, CRBM, in Section 3.
In Section 4, we present improvement methods for the CRBM using the CPCA.
In Section 5, we propose a complex-valued sequence generation method based on MLPG.
In Section 6, we show our experimental results, and we conclude our findings in Section 7. 

\begin{figure}[t]
  \centering
  \includegraphics[width=0.7\columnwidth]{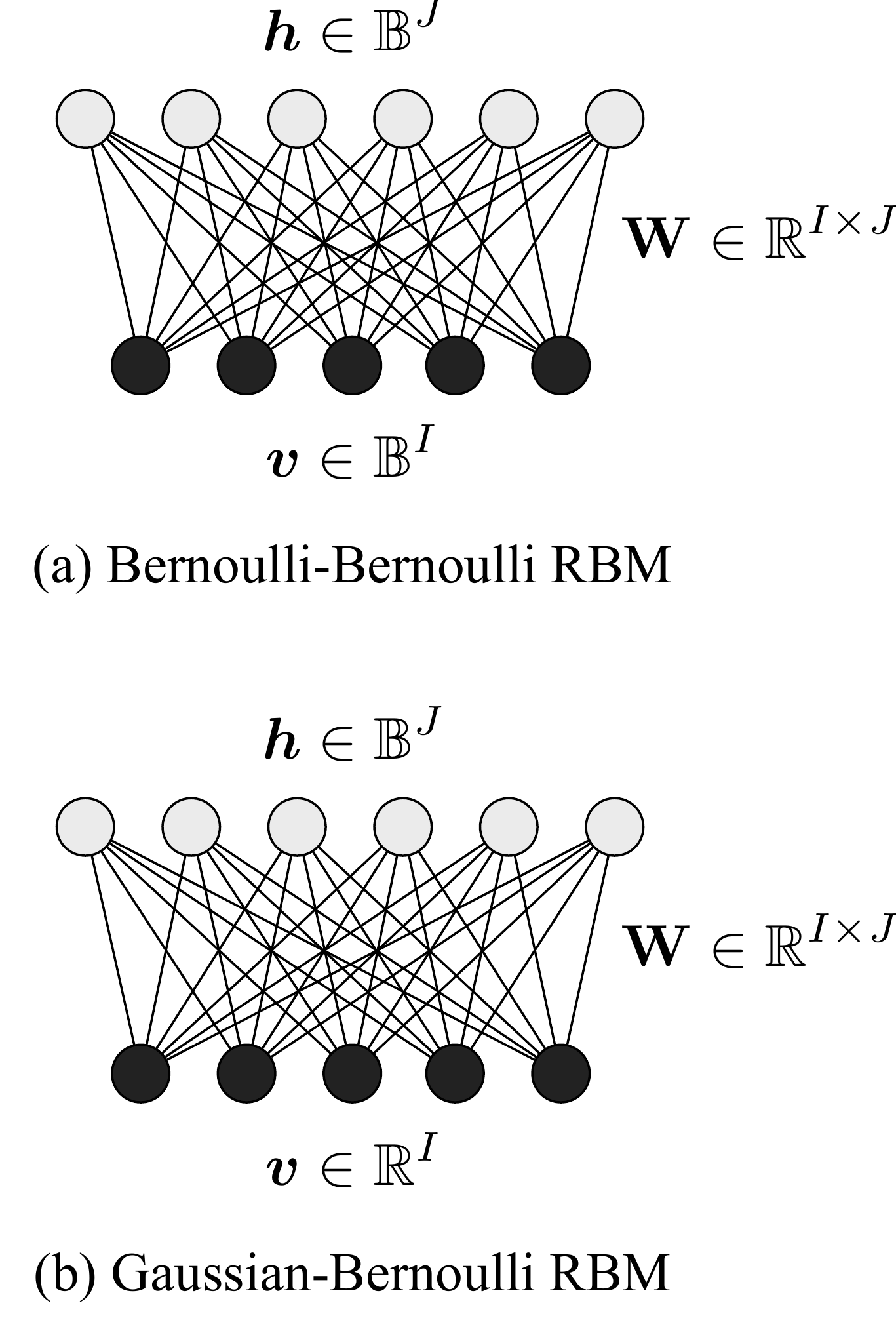}
  \caption{Graphical representation of conventional RBMs.}
  \label{fig:rbm}
\end{figure}

\section{Preliminary}

A restricted Boltzmann machine (RBM) is one of the most widely used energy-based models and is convenient for representing latent features that cannot be observed but surely exist in the background.
The Bernoulli-Bernoulli RBM (BB-RBM) RBM was originally introduced by Freund {\it et. al} \cite{Freund:1994tu}.
It defines the distribution of binary-valued visible variables $\bs{v} \in \mathbb{B}^I$ and binary-valued hidden (latent) variables $\bs{h} \in \mathbb{R}^J$ with their undirected real-valued connection weights $\mathbf{W} \in \mathbb{R}^{I \times J}$, as shown in Fig.~\ref{fig:rbm}~(a) where $I$ and $J$ are the numbers of dimensions in their respective visible and hidden units and $\mathbb{B} \triangleq \{0, 1\}$ indicates the binary set.
The RBM was later extended to deal with real-valued data known as a Gaussian-Bernoulli RBM (GB-RBM) \cite{Lee:2008uz}, as shown in Fig.~\ref{fig:rbm}~(b).
However, it has been reported that there were some difficulties with the original GB-RBM because of the unstable training of the parameters.
Later, Cho \textit{et al.} \cite{cho2011improved} proposed an improved learning method for a GB-RBM to overcome the difficulties.
In the remainder of this paper, we refer to this improved GB-RBM just as an RBM unless otherwise stated.
In the modeling using an RBM, the joint probability $p(\bs{v},\bs{h})$ of real-valued visible units $\bs{v}$ and binary-valued hidden units $\bs{h}$ is defined as follows:
\begin{align}
\label{eq:rbm_p}
p(\bs{v},\bs{h};\bs{\theta}) &= \frac{1}{U(\bs{\theta)}} e^{-E(\bs{v},\bs{h};\bs{\theta})} \\
\label{eq:rbm_e}
E(\bs{v},\bs{h};\bs{\theta}) &= \frac{1}{2} \bs{v}^\top \mathbf{\Sigma}^{-1} \bs{v} - \bs{b}^\top \mathbf{\Sigma}^{-1} \bs{v} - \bs{c}^\top \bs{h} - \bs{v}^\top \mathbf{\Sigma}^{-1} \mathbf{W} \bs{h} \\
\label{eq:rbm_u}
U(\bs{\theta}) &= \int \sum_{\bs{h}} e^{-E(\bs{v},\bs{h};\bs{\theta})} d \bs{v}
\end{align}
where $\bs{\theta} = \{\bs{b},\bs{c},\mathbf{W},\bs{\sigma}\}$ indicates a set of parameters that contains bias parameters of the visible units $\bs{b} \in \mathbb{R}^I$, bias parameters of the hidden units $\bs{c} \in \mathbb{R}^J$, the connection weight parameters between visible-hidden units $\mathbf{W} \in \mathbb{R}^{I \times J}$, and the standard deviation parameters associated with the dimension independent Gaussian visible units $\bs{\sigma} \in \mathbb{R}^I$ that define $\mathbf{\Sigma} \triangleq {\rm diag}(\bs{\sigma}^2$) (the function ${\rm diag}(\cdot)$ returns a diagonal matrix whose diagonal vector is the argument, and $\cdot^2$ indicates the element-wise square operation).
The parameters $\bs{\theta}$ are often estimated using the maximum likelihood (ML) and the gradient descent/ascent given the training set $D \ni \bs{v}$.
The partial gradients of the parameters to the expected log likelihood:
\begin{align}
L(\bs{\theta}) = \mathbb{E}_D [\log p(\bs{v}; \bs{\theta})] = \mathbb{E}_D [\log \sum_{\bs{h}} p(\bs{v}, \bs{h}; \bs{\theta})]
\end{align}
can be calculated as:
\begin{align}
\frac{\partial L}{\partial \bs{\theta}} = \langle -\frac{\partial E}{\partial \bs{\theta}} \rangle_{\rm data} - \langle - \frac{\partial E}{\partial \bs{\theta}} \rangle_{\rm model},
\end{align}
where $\langle \cdot \rangle_{\rm data}$ and $\langle \cdot \rangle_{\rm model}$ indicate expectations of the training data and the inner model, respectively.
Although exact calculation of the inner model has an order of $2^{I + J}$, the expectation value can be approximated using the Gibbs sampling, or more efficiently, the contrastive divergence (CD) \cite{hinton2006fast}.
From the definition of RBM in Eqs. \eqref{eq:rbm_p}, \eqref{eq:rbm_e}, and \eqref{eq:rbm_u}, the conditional probabilities $p(\bs{v}|\bs{h})$ and $p(\bs{h}|\bs{v})$ form quite simple distributions as:
\begin{align}
\label{eq:rbm_pvh}
p(\bs{v}|\bs{h}) &= \mathcal{N}(\bs{v}; \bs{b} + \mathbf{W} \bs{h}, \mathbf{\Sigma}) \\
\label{eq:rbm_phv}
p(\bs{h}|\bs{v}) &= \mathcal{B}(\bs{h}; \bs{f}(\bs{c} + \mathbf{W}^\top \mathbf{\Sigma}^{-1} \bs{v}))
\end{align}
where $\mathcal{N}(\cdot; \bs{\mu}, \mathbf{\Sigma})$, $\mathcal{B}(\cdot; \bs{\pi})$, and $\bs{f}(\cdot)$ indicate the multivariate Gaussian distribution with the mean $\bs{\mu}$ and the variance matrix $\mathbf{\Sigma}$, the multi-dimensional Bernoulli distribution with the success probabilities $\bs{\pi}$, and an element-wise sigmoid function, respectively.
As Eqs.~\eqref{eq:rbm_pvh} and \eqref{eq:rbm_phv} indicates, we can easily compute the iteration of drawing samples $\bs{h}$ given $\bs{v}$, and $\bs{v}$ given $\bs{h}$, which is used in Gibbs sampling or CD.
The same is true for BB-RBM.
In the case of BB-RBM, the conditional probabilities $p(\bs{v}|\bs{h})$ and $p(\bs{h}|\bs{v})$ turn into the following:
\begin{align}
\label{eq:bbrbm_pvh}
p(\bs{v}|\bs{h}) &= \mathcal{B}(\bs{v}; \bs{f}(\bs{b} + \mathbf{W} \bs{h}) \\
p(\bs{h}|\bs{v}) &= \mathcal{B}(\bs{h}; \bs{f}(\bs{c} + \mathbf{W}^\top \bs{v}))
\end{align}
under the energy function:
\begin{align}
E(\bs{v},\bs{h};\bs{\theta}) = - \bs{b}^\top \bs{v} - \bs{c}^\top \bs{h} - \bs{v}^\top \mathbf{W} \bs{h}.
\end{align}

\begin{figure}[t]
  \centering
  \includegraphics[width=0.85\columnwidth]{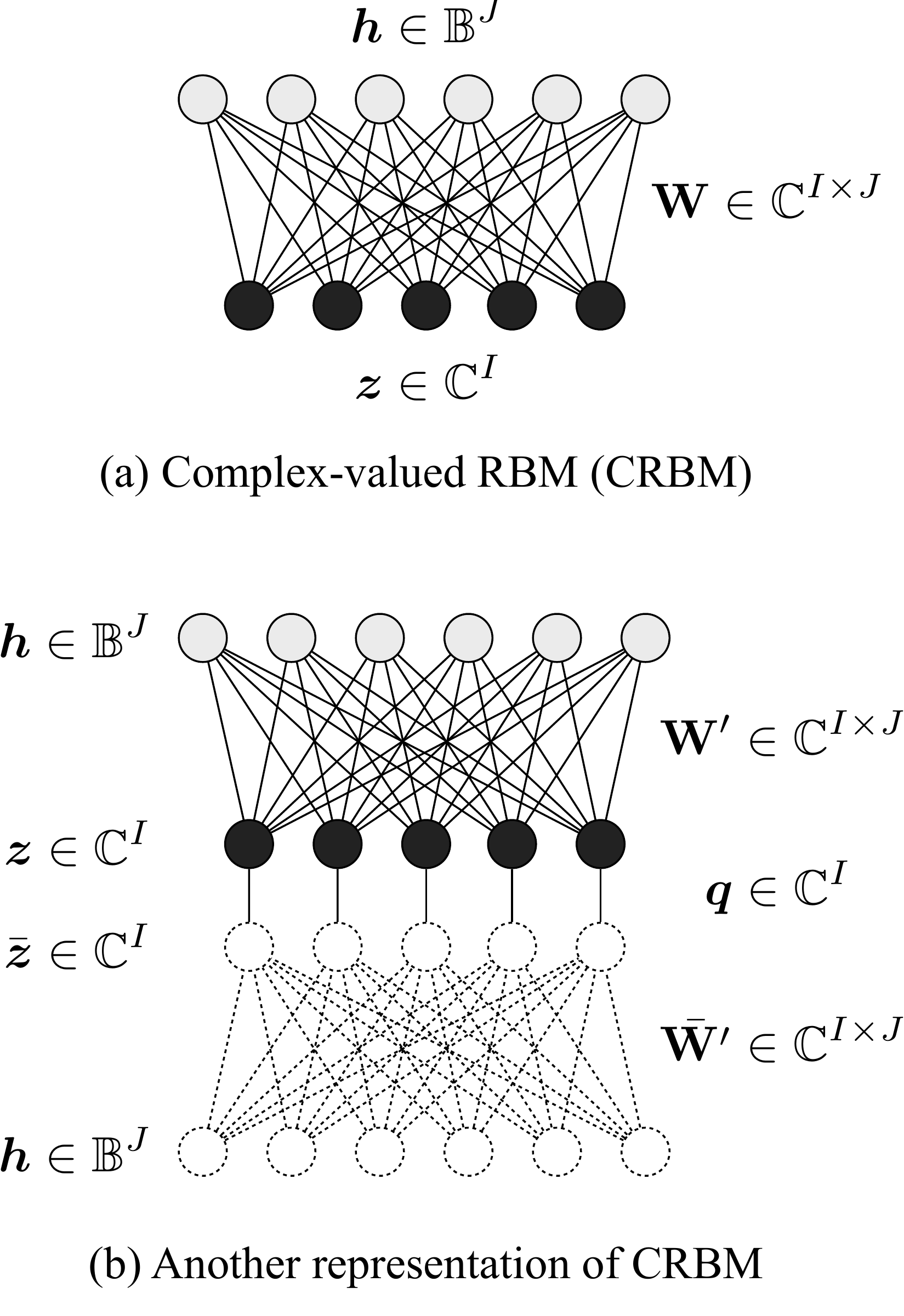}
  \caption{Graphical representation of a complex-valued RBM.}
  \label{fig:crbm}
\end{figure}

\section{Complex-valued RBM}

Conventional RBMs assume that data is either binary-valued or real-valued.
Therefore, complex-valued data should not be fed into the conventional RBMs directly because the conditional probability of visible units specifies binary- or real-valued variables, as Eqs.~\eqref{eq:rbm_pvh} and \eqref{eq:bbrbm_pvh} indicate.
In other words, the conditional probability of visible units should specify complex-valued variables in order to feed complex-valued variables into the model.
In our approach, we define an extension of the RBM that feeds complex-valued data and forms the conditional probability of visible units as complex normal distribution \cite{picinbono1996second}.
A real-valued cost function (the likelihood) is still used in parameter estimation for an extended RBM---namely a complex-valued RBM (CRBM---because the probability distribution is real-valued.
Like conventional RBMs, the CRBM consists of two layers: complex-valued visible units $\bs{z}$ and binary-valued hidden units $\bs{h}$ with undirected connection weights $\mathbf{W}$, as shown in Fig.~\ref{fig:crbm}~(a).
Furthermore, in the CRBM, wegive a ``restriction'' where there are no connections between visible units or hidden units, which enables easy estimation of parameters just as an RBM does.
However, we allow the model to have connections between the real and imaginary parts in order to capture the relationships between the real and imaginary parts of each complex-valued visible unit.

\subsection{Definition}

Based on the above discussion, we formulated a CRBM that has $I$-dimensional complex-valued visible units $\bs{z} \in \mathbb{C}^I$ and $J$-dimensional binary-valued hidden units $\bs{h} \in \mathbb{B}^J$ as follows:
\begin{align}
p(\bs{z},\bs{h};\bs{\theta}) =& \frac{1}{U(\bs{\theta)}} e^{-E(\bs{z},\bs{h};\bs{\theta})} \\
\begin{split}
\label{eq:energy} E(\bs{z},\bs{h};\bs{\theta}) =& \frac{1}{2} \left[ \begin{array}{l} \bs{z} \\ \bar{\bs{z}} \end{array} \right]^H \mathbf{\Phi}^{-1} \left[ \begin{array}{l} \bs{z} \\ \bar{\bs{z}} \end{array} \right] \\ &- \left[ \begin{array}{l} \bs{b} \\ \bar{\bs{b}} \end{array} \right]^H \mathbf{\Phi}^{-1} \left[ \begin{array}{l} \bs{z} \\ \bar{\bs{z}} \end{array} \right] - 2 \bs{c}^\top \bs{h} \\ &- \left[ \begin{array}{l} \bs{z} \\ \bar{\bs{z}} \end{array} \right]^H \mathbf{\Phi}^{-1} \left[ \begin{array}{l} \mathbf{W} \\ \bar{\mathbf{W}} \end{array} \right] \bs{h}
\end{split} \\
U(\bs{\theta}) =& \int \sum_{\bs{h}} e^{-E(\bs{z},\bs{h};\bs{\theta})} d \bs{z},
\end{align}
where $\bar{\cdot}$ denotes complex-conjugate and $\cdot^H$ denotes Hermitian-transpose.
$\bs{b} \in \mathbb{C}^I$, $\bs{c} \in \mathbb{R}^J$, and $\mathbf{W} \in \mathbb{C}^{I \times J}$ are bias parameters of the visible units and the hidden units, and the \textit{biased} connection weights between visible and hidden units, respectively.
In order to make the restrictions, the extended covariance matrix $\mathbf{\Phi}$ consists of a covariance matrix $\mathbf{\Gamma}$ and a pseudo-covariance matrix $\mathbf{C}$---both of which are diagonal matrices---as
\begin{equation}
\mathbf{\Phi} \triangleq \left[ \begin{array}{ll} \mathbf{\Gamma} & \mathbf{C} \\ \mathbf{C}^H & \mathbf{\Gamma}^H \end{array} \right]
\end{equation}
and
\begin{align}
\begin{array}{ll}
\mathbf{\Gamma} \triangleq {\rm diag}(\bs{\gamma}), & \bs{\gamma} \in \mathbb{R+}^I \\
\mathbf{C} \triangleq {\rm diag}(\bs{\delta}), & \bs{\delta} \in \mathbb{C}^I
\end{array}
\end{align}
where $\bs{\gamma}$ and $\bs{\delta}$ are variance and pseudo-variance parameters of the complex-valued visible units, respectively.
To summarize, the set of parameters of the CRBM is $\bs{\theta} = \{\bs{b},\bs{c},\mathbf{W},\bs{\gamma},\bs{\delta} \}$.

Introducing auxiliary precision vectors $\bs{p}$ and $\bs{q}$ defined as
\begin{align}
\bs{p} &\triangleq \frac{\bs{\gamma}}{\bs{\gamma}^2 - |\bs{\delta}|^2} \in \mathbb{R}^I \\
\bs{q} &\triangleq - \frac{\bs{\delta}}{\bs{\gamma}^2 - |\bs{\delta}|^2} \in \mathbb{C}^I
\end{align}
where the fraction bar denotes element-wise division, we can rewrite the energy function in Eq.~\eqref{eq:energy} as follows:
\begin{align}
\begin{split}
&E(\bs{z},\bs{h};\bs{\theta}) = \\
&\bs{z}^H {\rm diag}(\bs{p}) \bs{z} + \Re(\bs{z}^H {\rm diag}(\bs{q})\bar{\bs{z}}) - 2\Re(\bs{z}^H {\rm diag}(\bs{p}) \bs{b}) \\ &-2\Re(\bs{z}^H {\rm diag}(\bs{q}) \bar{\bs{b}}) - 2 \bs{c}^\top \bs{h} -2\Re(\bs{z}^H {\rm diag}(\bs{p})\mathbf{W}) \bs{h} \\ &-2\Re(\bs{z}^H {\rm diag}(\bs{q})\bar{\mathbf{W}}) \bs{h},
\end{split}
\end{align}
which confirms that 1) the above energy function $E$ and the probability distribution are real-valued and that 2) there are connections between the complex-valued visible units and their conjugates for each dimension but there are no connections between different dimensions.

Furthermore, when we use unbiased parameters:
\begin{align}
\bs{b}' &\triangleq {\rm diag}(\bs{p}) \bs{b} + {\rm diag}(\bs{q}) \bar{\bs{b}} \\
\mathbf{W}' &\triangleq {\rm diag}(\bs{p}) \mathbf{W} + {\rm diag}(\bs{q}) \bar{\mathbf{W}},
\end{align}
the energy function $E$ becomes
\begin{align}
\begin{split}
&E(\bs{z},\bs{h};\bs{\theta}) = \\
&\frac{1}{2} \bs{z}^H {\rm diag}(\bs{p}) \bs{z} + \frac{1}{2} \bar{\bs{z}}^H {\rm diag}(\bs{p}) \bar{\bs{z}} + \bs{z}^H {\rm diag}(\bs{q})\bar{\bs{z}} \\ &+ \bar{\bs{z}}^H {\rm diag}(\bar{\bs{q}}) \bs{z} - \bs{z}^H \bs{b}' - \bar{\bs{z}}^H \bar{\bs{b}}' - 2 \bs{c}^\top \bs{h} \\ &- \bs{z}^H \mathbf{W}' \bs{h} - \bar{\bs{z}}^H \bar{\mathbf{W}}' \bs{h},	
\end{split}
\end{align}
which indicates that $\bs{z}$ and $\bar{\bs{z}}$ are symmetric to each other, as shown in Figure~\ref{fig:crbm}~(b).

From the above definition, the conditional probabilities $p(\bs{z}|\bs{h})$ and $p(\bs{h}|\bs{z})$ can be derived as follows:
\begin{align}
\label{eq:crbm_pzh}
p(\bs{z}|\bs{h}) &= \mathcal{CN}(\bs{z}; \bs{b} + \mathbf{W} \bs{h}, \mathbf{\Gamma}, \mathbf{C}) \\
\label{eq:crbm_phz}
p(\bs{h}|\bs{z}) &= \mathcal{B}(\bs{h}; \bs{f}(2 \bs{c} + 2 \Re (\mathbf{W}'^H \bs{z} ) ))
\end{align}
where $\mathcal{CN}(\cdot; \bs{\mu}, \mathbf{\Gamma}, \mathbf{C})$ is a multivariate complex normal distribution \cite{picinbono1996second} a mean vector $\bs{\mu}$, a covariance matrix $\mathbf{\Gamma}$, and a pseudo-covariance matrix $\mathbf{C}$:
\begin{align}
\begin{split}
&p(\bs{z}) = \frac{1}{\pi^D \sqrt{\rm det(\mathbf{\Gamma)} \rm det(\mathbf{Q})}} \\ &\cdot \exp \left\{ -\frac{1}{2} \left[ \begin{array}{l} \bs{z}-\bs{\mu} \\ \bar{\bs{z}} - \bar{\bs{\mu}} \end{array} \right]^H \left[ \begin{array}{ll} \mathbf{\Gamma} & \mathbf{C} \\ \mathbf{C}^H & \mathbf{\Gamma}^H \end{array} \right]^{-1} \left[ \begin{array}{l} \bs{z}-\bs{\mu} \\ \bar{\bs{z}} - \bar{\bs{\mu}} \end{array} \right] \right\}
\end{split} \\
&\mathbf{Q} = \bar{\mathbf{\Gamma}} - \mathbf{C}^H \mathbf{\Gamma}^{-1} \mathbf{C}.
\end{align}

\subsection{Parameter estimation}

To estimate the model parameters $\bs{\theta}$ of the CRBM, we employed the complex-valued gradient method.
In this approach, the parameters $\bs{\theta}$ are estimated so as to maximize the expected log-likelihood $L$ of the complex-valued training data set $D \ni \bs{z}$:
\begin{align}
\label{eq:L} L(\bs{\theta}) &= \mathbb{E}_D [\log p(\bs{z};\bs{\theta}) ]\\
&= \mathbb{E}_D [ \log \sum_{\bs{h}} p(\bs{z},\bs{h};\bs{\theta}) ]\\
&= \mathbb{E}_D [\log \sum_{\bs{h}} e^{-E(\bs{z},\bs{h};\bs{\theta})}] - \log \int \sum_{\tilde{\bs{h}}} e^{-E(\tilde{\bs{z}},\tilde{\bs{h}};\bs{\theta})} d\tilde{\bs{z}}.
\end{align}
The complex-valued gradient ascend iteratively updates each parameter as:
\begin{align}
\label{eq:update}
\bs{\theta}^{(l+1)} \gets \bs{\theta}^{(l)} + \bs{g}^{(l)} (\frac{\partial L}{\partial \bs{\theta}})
\end{align}
where $\bs{\theta}^{(l)}$ indicates the parameters and $\bs{g}^{(l)}$ indicates the complex-valued gradient at the $l$-th iteration.
One of the simplest gradient functions is the complex-valued steepest ascent (CSA) \cite{Brandwood:1983ds,Zhang:2016gd}, which is:
\begin{align}
\bs{g}^{(l)} (\frac{\partial L}{\partial \bs{\theta}}) = 2 \alpha \frac{\partial L}{\partial \bs{\bar{\theta}}}
\end{align}
where $\alpha \in \mathbb{C}, \Re(\alpha)>0$ is a complex-valued learning rate.
A simple CSA is not be suitable for a large amount of training data of speech due to the slow convergence speed.
Therefore, we propose another, more efficient learning method, the complex-valued adaptive momentum (CAdam), which is motivated by the real-valued Adam algorithm \cite{kingma2015adam}.
In the CAdam, we introduce auxiliary parameters $\bs{m}^{(n)}$ and $\bs{v}^{(n)}$ and update the parameters as:
\begin{align}
\bs{m}^{(l)} &= \beta_1 \bs{m}^{(n-1)} + (1-\beta_1) \nabla_{\bar{\bs{\theta}}} L \\
\bs{v}^{(l)} &= \beta_2 \bs{v}^{(n-1)} + (1-\beta_2) | \nabla_{\bar{\bs{\theta}}} L |^2 \\
\Delta \bs{\theta}^{(l)} &= 2 \alpha \frac{1-\beta_2^l}{1-\beta_1^l} \frac{\bs{m}^{(l)}}{\bs{v}^{(l)}},
\end{align}
where $\beta_1, \beta_2 \in \mathbb{C}, 0 < |\beta_1|, |\beta_2| < 1$, and $\alpha \in \mathbb{C}, \Re(\alpha)>0$.

Calculating partial gradients to the parameters $\bs{\theta}$, we obtain:
\begin{align}
\label{eq:crbm_dl}
\frac{\partial L}{\partial \bs{\theta}} = \langle - \frac{\partial E}{\partial \bs{\theta}} \rangle_{\rm data} - \langle - \frac{\partial E}{\partial \bs{\theta}} \rangle_{\rm model},
\end{align}
where the complex-valued partial gradients here indicate the Wirtinger derivatives:
\begin{align}
\frac{\partial L}{\partial \bs{\theta}} &= \frac{1}{2} \left( \frac{\partial L}{\partial \Re(\bs{\theta})} - i \frac{\partial L}{\partial \Im(\bs{\theta})} \right) \\
\frac{\partial L}{\partial \bar{\bs{\theta}}} &= \frac{1}{2} \left( \frac{\partial L}{\partial \Re(\bs{\theta})} + i \frac{\partial L}{\partial \Im(\bs{\theta})} \right).
\end{align}
The negative partial gradients of the energy function with respect to each parameter $- \frac{\partial E}{\partial \bs{\theta}}$ can be further derived as:
\begin{align}
\label{eq:b} - \frac{\partial E}{\partial \bs{b}} &= {\rm diag}(\bs{p}) \bar{\bs{z}} + {\rm diag}(\bar{\bs{q}}) \bs{z} \\
- \frac{\partial E}{\partial \bs{c}} &= \bs{h} \\
- \frac{\partial E}{\partial \mathbf{W}} &= ({\rm diag}(\bs{p}) \bar{\bs{z}} + {\rm diag}(\bar{\bs{q}}) \bs{z}) \bs{h}^\top \\
- \frac{\partial E}{\partial \bs{\gamma}} &= (\bs{p}^2 + |\bs{q}|^2) \circ \frac{\partial E}{\partial \bs{p}} + 2 \Re(\bs{p} \circ \bs{q} \circ \frac{\partial E}{\partial \bs{q}}) \\
- \frac{\partial E}{\partial \bs{\delta}} &= \bs{p}^2 \circ \frac{\partial E}{\partial \bs{q}} + \bar{\bs{q}}^2 \circ \frac{\partial E}{\partial \bar{\bs{q}}} + 2 \bs{p} \circ \bar{\bs{q}} \circ \frac{\partial E}{\partial \bs{p}},
\end{align}
where $\circ$ denotes an element-wise product and $|\cdot|$ denotes the absolute, and
\begin{align}
\frac{\partial E}{\partial \bs{p}} &= \frac{1}{2} |\bs{z} |^2 - \Re(\bs{z} \circ (\bar{\bs{b}} + \bar{\mathbf{W}} \bs{h})) \\
\frac{\partial E}{\partial \bs{q}} &= \frac{1}{2} \bar{\bs{z}}^2 - \bar{\bs{z}} \circ (\bar{\bs{b}} + \bar{\mathbf{W}} \bs{h}).
\end{align}
The gradients of variance and pseudo variance tend to be larger than those of the other parameters.
For stable training, we replace the parameters as $\bs{\gamma} \triangleq e^{\bs{r}}$ and $\bs{\delta} \triangleq e^{\bs{s}}$ and update using the gradients of $\bs{r}$ and $\bs{s}$ in a manner similar to the improved GB-RBM \cite{cho2011improved}.

The second term on the right-hand side in Eq.~\eqref{eq:crbm_dl} usually requires high computational cost.
However, because of the restrictions of the CRBM, the second term can be efficiently approximated using  Gibbs sampling or CD \cite{hinton2006fast} in a way similar to conventional RBMs.

\subsection{Relationships with complex representation using GB-RBM}

\begin{figure}[t]
  \centering
  \includegraphics[width=1.00\columnwidth]{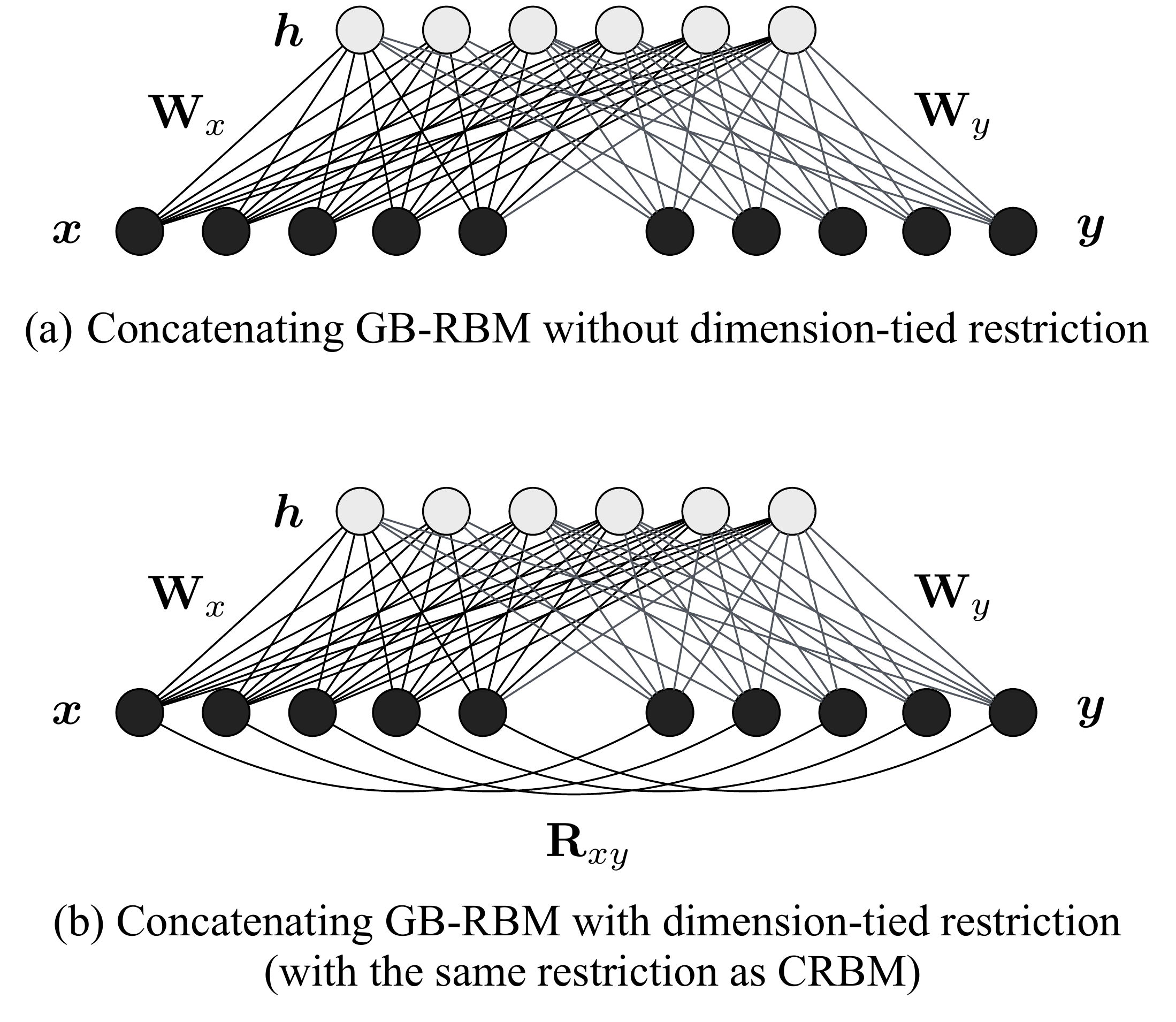}
  \caption{Representing complex-valued data using concatenating GB-RBM (a) without and (b) with dimension-tied restriction.}
  \label{fig:conrbm}
\end{figure}

We can also represent a complex-valued vector $\bs{z} = \bs{x} + i \bs{y}$ (where $\bs{x}, \bs{y} \in \mathbb{R}^I$) in the real space using a conventional GB-RBM that feeds the double-sized concatenated vector $\bs{z}' = [ \bs{x}^\top \ \bs{y}^\top ]^\top \in \mathbb{R}^{2I}$ as:
\begin{align}
p(\bs{x},\bs{y}) &= \sum	_{\bs{h}} p(\bs{x},\bs{y},\bs{h}) \\
p(\bs{x},\bs{y},\bs{h}) &= \frac{1}{U} e^{-E(\bs{x},\bs{y},\bs{h})} \\
\label{eq:econ1} E(\bs{x},\bs{y},\bs{h}) &= \frac{1}{2} \bs{x}^\top \mathbf{\Sigma}^{-1}_x \bs{x} + \frac{1}{2} \bs{y}^\top \mathbf{\Sigma}^{-1}_y \bs{y} \nonumber \\ & \ \ \ - \bs{b}_x^\top \mathbf{\Sigma}^{-1}_x \bs{x} - \bs{b}_y^\top \mathbf{\Sigma}^{-1}_y \bs{y} - \bs{c}^\top \bs{h} \\ & \ \ \ - \bs{x}^\top \mathbf{\Sigma}^{-1}_x \mathbf{W}_x \bs{h} - \bs{y}^\top \mathbf{\Sigma}^{-1}_y \mathbf{W}_y \bs{h} \nonumber \\
U &= \int \int \sum_{\bs{h}} e^{-E(\bs{x},\bs{y},\bs{h})} d\bs{x} d\bs{y},
\end{align}
where $\mathbf{\Sigma}_x = {\rm diag} (\bs{\sigma}_x^2), \mathbf{\Sigma}_y = {\rm diag} (\bs{\sigma}_y^2)$ and we decompose the GB-RBM parameters as $\bs{b} = [ \bs{b}_x^\top \ \bs{b}_y^\top ]^\top$, $\bs{\sigma} = [ \bs{\sigma}_x^\top \ \bs{\sigma}_y^\top ]^\top$, $\mathbf{W} = [ \mathbf{W}_x^\top \ \mathbf{W}_y^\top ]^\top$ in Eqs. \eqref{eq:rbm_p}, \eqref{eq:rbm_e}, and \eqref{eq:rbm_u}.
Fig.~\ref{fig:conrbm}~(a) depicts this concatenating representation of the complex-valued data using GB-RBM.
For example, the negative partial differentials of the real and imaginary parts of the bias parameters to the energy function in this representation can be derived as:
\begin{align}
\label{eq:dedbx} - \frac{\partial E}{\partial \bs{b}_x} &= \frac{\bs{x}}{\bs{\sigma}_x^2} \\
\label{eq:dedby} - \frac{\partial E}{\partial \bs{b}_y} &= \frac{\bs{y}}{\bs{\sigma}_y^2}.
\end{align}

On the other hand, when we put $\bs{z} = \bs{x} + i \bs{y}$, $\bs{b} = \bs{b}^R + i \bs{b}^I$, $\mathbf{W} = \mathbf{W}^R + i \mathbf{W}^I$, $\bs{q} = \bs{q}^R + i \bs{q}^I$ where $\bs{b}^R, \bs{b}^I \in \mathbb{R}^I$, $\mathbf{W}^R, \mathbf{W}^I \in \mathbb{R}^{I \times J}$, $\bs{q}^R, \bs{q}^I \in \mathbb{R}^I$, we can rewrite the energy function of CRBM in Eq.~\eqref{eq:energy} as:
\begin{align}
\label{eq:econ2}
\begin{split} E = & \frac{1}{2} \bs{x}^\top \mathbf{\Sigma}^{-1}_x \bs{x} + \bs{x}^\top \mathbf{R}_{xy} \bs{y} + \frac{1}{2} \bs{y}^\top \mathbf{\Sigma}^{-1}_y \bs{y} \\ &- \bs{b}_x^\top \mathbf{\Sigma}^{-1}_x \bs{x} - \bs{b}_y^\top \mathbf{\Sigma}^{-1}_y \bs{y} - \bs{c}^\top \bs{h} \\ &- \bs{x}^\top \mathbf{\Sigma}^{-1}_x \mathbf{W}_x \bs{h} - \bs{y}^\top \mathbf{\Sigma}^{-1}_y \mathbf{W}_y \bs{h} \end{split},
\end{align}
where we introduce
\begin{align}
\mathbf{\Sigma}_x &= {\rm diag}(\frac{1}{2(\bs{p} + \bs{q}^R)}) \\
\mathbf{\Sigma}_y &= {\rm diag}(\frac{1}{2(\bs{p} - \bs{q}^R)}) \\
\mathbf{R}_{xy} &= {\rm diag}(2\bs{q}^I) \\
\bs{b}_x &= \bs{b}^R + \frac{\bs{q}^I}{\bs{p}+\bs{q}^R} \circ \bs{b}^I \\
\bs{b}_y &= \bs{b}^I + \frac{\bs{q}^I}{\bs{p}-\bs{q}^R} \circ \bs{b}^R \\
\mathbf{W}_x &= \mathbf{W}^R + {\rm diag}(\frac{\bs{q}^I}{\bs{p}+\bs{q}^R}) \mathbf{W}^I \\
\mathbf{W}_y &= \mathbf{W}^I + {\rm diag}(\frac{\bs{q}^I}{\bs{p}-\bs{q}^R}) \mathbf{W}^R.
\end{align}

Comparing the energy functions in Eqs.~\eqref{eq:econ1} and \eqref{eq:econ2}, the latter energy function includes a cross term ($\bs{x}^\top \mathbf{R}_{xy} \bs{y}$) between $\bs{x}$ and $\bs{y}$ while the former energy function does not.
Therefore, we can claim that the CRBM representation extends the conventional GB-RBM, which has connections between the real and imaginary parts for each dimension with the weights $\bs{r}_{xy} = 2 \bs{q}^I$ as in Fig.~\ref{fig:conrbm}~(b), where $\bs{r}_{xy}$ is the diagonal vector of $\mathbf{R}_{xy}$.

Furthermore, in the GB-RBM representation, the gradients regarding the real and imaginary parts of the bias of visible units, for example, are calculated independently of each other as Eqs.~\eqref{eq:dedbx} and \eqref{eq:dedby} indicate, while the gradients of the bias of visible units in the CRBM representation are calculated using both the real and imaginary terms as Eq.~\eqref{eq:b} indicates.
This will make the model convergence better than the GB-RBM.

\section{Complex spectra compression using CPCA}
\label{sec:cpca}

The aim of this paper is to represent trajectories of complex-valued speech spectra.
In general, the number of dimensions of the raw complex spectra tends to be large (e.g., when analyzing speech with the window length of $1,024$, the complex spectra has the dimensions of $513$), which makes it difficult to use dynamic features or segment features as input for the model due to the sizable number of parameters.
Therefore, we reduced the dimensions using the complex principal component analysis (CPCA) \cite{horel1984complex} in this paper.

Letting $\bs{o}_t$ be the complex spectra at the frame $t$, the complex-valued features $\bs{z}_t$ whose dimensions are reduced to $P$ using CPCA calculated as:
\begin{align}
\bs{z}_t = \mathbf{\Lambda}_{1:P}^{-\frac{1}{2}} \mathbf{U}_{:,1:P}^H \bs{o}_t,
\end{align}
where $\mathbf{\Lambda}_{1:P}^{-\frac{1}{2}}$ and $\mathbf{U}_{:,1:P}$ indicate a diagonal matrix where the diagonal elements are the inverse of the top $P$ eigenvalues of the empirical covariance matrix and a complex matrix whose columns are the complex eigenvectors corresponding to the eigenvalues.
Conversely, when we recover the complex spectra $\bs{o}_t$ from $\bs{z}_t$, we just calculate the inversion as:
\begin{align}
\label{eq:ipca}
\bs{o}_t = \mathbf{U}_{:,1:P} \mathbf{\Lambda}_{1:P}^{\frac{1}{2}} \bs{z}_t.
\end{align}

In our speech modeling experiments that will be discussed later, we used the concatenated features $\bs{Z}_t \triangleq [\bs{z}_t^H \ \Delta \bs{z}_t^H]^H$ as visible units in  CRBM, where $\bs{z}_t^H$ was calculated using the CPCA with the degree of $P=40$ from the complex spectra analyzed with the window length of $256$, and their dynamics $\Delta \bs{z}_t^H$ were calculated as $0.5\bs{z}_{t+1} - 0.5\bs{z}_{t-1}$.
The total dimensions of visible units in CRBM were $I=80$.
In these experiments, the CRBM was trained so as to maximize the expected likelihood $L(\bs{\theta}) = \mathbb{E}[\log p(\bs{Z};\bs{\theta})]$ of the concatenated feature set.

\section{Complex-valued sequence generation based on MLPG}

When we apply the CRBM to represent speech spectra, we need further improvements to compare it to other feature extraction methods of speech.
In this section, improved methods of dealing with trajectory modeling will be presented.

In our first work on the CRBM \cite{nakashika2017complex}, complex-valued visible units $\bs{z}_t$ were probabilistically encoded into binary values by calculating the expectations of hidden units as $\hat{\bs{h}}_t \triangleq \mathbb{E}[p(\bs{h}_t|\bs{z}_t)]$ and inversely decoded (recovered) from $\hat{\bs{h}}_t$ by calculating the expectations of visible units as $\hat{\bs{z}}_t \triangleq \mathbb{E}[p(\bs{z}_t|\hat{\bs{h}}_t)] = \bs{b} + \mathbf{W} \hat{\bs{h}}_t$ frame-by-frame.
However, speech signals are sequences; there are correlations between adjacent frames of speech.
In this paper, we employ trajectory modeling and sequence generation instead of frame-wise modeling.
The proposed, efficient method recovers complex-valued visible units involving correlations among the neighbor frames based on the maximum likelihood parameter generation (MLPG) \cite{tokuda2000speech}.
The MLPG is the algorithm to estimate the optimum sequence of features from static and dynamic features based on a maximum likelihood estimation.
This cannot be directly applied to complex-valued features; therefore, we present the following formulation.

After training the CRBM, we estimated the optimum sequence of CPCA features $\hat{\bs{z}}_{1:T} \triangleq [\hat{\bs{z}}_1^H \ \hat{\bs{z}}_2^H \ \cdots \ \hat{\bs{z}}_T^H]^H$, where $T$ is the number of frames of the test speech, from the encoded features (the expectations of hidden units) $\hat{\bs{h}}_t \triangleq \mathbb{E}[p(\bs{h}_t|\bs{z}_t)]$ that were calculated from the original concatenated features $\bs{Z}_{1:T} \triangleq [\bs{Z}_1^H \ \bs{Z}_2^H \ \cdots \ \bs{Z}_T^H]^H$ of the test speech.
$\hat{\bs{z}}_{1:T}$ is the sequence that maximizes the conditional probability $p(\bs{Z}_{1:T} | \hat{\bs{h}}_{1:T})$, which is defined as:
\begin{align}
\hat{\bs{z}}_{1:T} = \argmax_{\bs{z}_{1:T}} p(\bs{Z}_{1:T} | \hat{\bs{h}}_{1:T}).
\end{align}
Now introducing the weight matrix $\mathbf{S} \in \mathbb{R}^{IT \times PT}$ that is:
\begin{align}
\mathbf{S} &\triangleq [\mathbf{S}_1 \ \mathbf{S}_2 \ \cdots \mathbf{S}_T]^\top \otimes \bs{I}_{P \times P} \\
\mathbf{S}_t &\triangleq [\bs{s}_t^{(1)} \ \bs{s}_t^{(2)}],
\end{align}
where $\bs{s}_t^{(1)} \in \mathbb{R}^{T}$ and $\bs{s}_t^{(2)} \in \mathbb{R}^{T}$ are the sparse vectors where only the $t$-th element is 1 otherwise 0, and where the $(t-1)$-th element has the value of -0.5 and the $(t+1)$-th element of 0.5 otherwise 0, respectively, the sequence can be rewritten as $\bs{Z}_{1:T} = \mathbf{S} \bs{z}_{1:T}$.
Since the conditional probability in Eq.\eqref{eq:crbm_pzh} has a single mode, the objective $\mathcal{Q} \triangleq \log p(\bs{Z}_{1:T} | \hat{\bs{h}}_{1:T}, \bs{\theta})$ can be calculated as:
\begin{align}
\begin{split}
Q = &- \bs{z}_{1:T}^\top \bs{S}^\top {\rm diag}(\tilde{\bar{\bs{q}}}) \bs{S} \bs{z}_{1:T} - \bs{z}_{1:T}^\top \bs{S}^\top {\rm diag}(\tilde{\bs{p}}) \bs{S} \bar{\bs{z}}_{1:T} \\
& + \bs{z}_{1:T}^\top \bs{S}^\top \bs{\mu}_{1:T} + K
\end{split},
\end{align}
where $K$ is a constant that can be ignored in the estimation, $\tilde{\bs{x}}$ indicates a vector that put $\bs{x}$ for $T$ times in a column, and
\begin{align}
\bs{\mu}_{1:T} \triangleq & \ [\bs{\mu}_1^H \ \bs{\mu}_2^H \ \cdots \ \bs{\mu}_T^H]^H \\
\bs{\mu}_t \triangleq & \ {\rm diag}(\bs{p}) (\bs{b} + \mathbf{W} \hat{\bs{h}}_t) + {\rm diag}(\bs{q}) (\bar{\bs{b}} + \bar{\mathbf{W}} \hat{\bs{h}}_t).
\end{align}
In this paper, we estimate the optimum sequence $\hat{\bs{z}}_{1:T}$ using a complex-valued gradient method in a way similar to that discussed in the previous section.
Specifically, using the initial sequence as the frame-wise optima of the static features from:
\begin{align}
\argmax_{\bs{Z}_t} p(\bs{Z}_t|\hat{\bs{h}}_t, \bs{\theta}) = \bs{b} + \mathbf{W} \hat{\bs{h}}_t, \forall t,
\end{align}
we iteratively update the sequence as:
\begin{align}
\label{eq:update_z}
\bs{z}_{1:T}^{(l+1)} \gets \bs{z}_{1:T}^{(l)} + \bs{g}^{(l)} (\frac{\partial Q}{\partial \bs{z}_{1:T}}),
\end{align}
where $\frac{\partial Q}{\partial \bs{z}_{1:T}}$ indicates the Wirtinger derivative and can be calculated as:
\begin{equation}
\frac{\partial Q}{\partial \bs{z}_{1:T}} = - 2 \bs{S}^\top {\rm diag}(\tilde{\bar{\bs{q}}}) \bs{S} \bs{z}_{1:T} - \bs{S}^\top {\rm diag}(\tilde{\bs{p}}) \bs{S} \bar{\bs{z}}_{1:T} + \bs{S}^\top \tilde{\bar{\bs{\mu}}}.
\end{equation}
In our experiments, we employ the CSA for the gradient function $\bs{g}^{(l)}$ in the sequence generation.

\section{Experimental evaluation}
\label{sec:expr}

\begin{figure}[t]
  \centering
  \includegraphics[width=0.8\columnwidth]{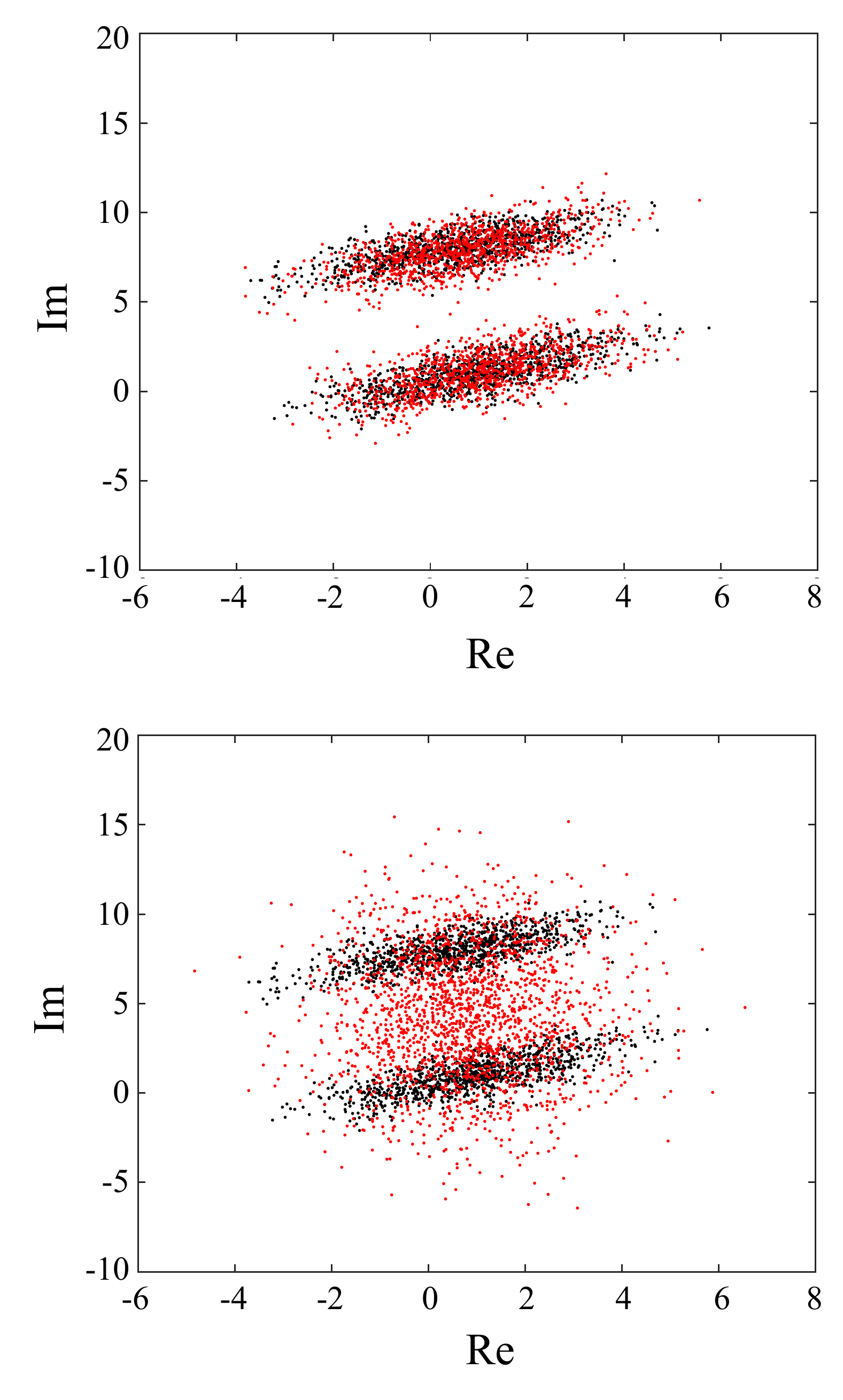}
  \caption{Artificially created 1D complex-valued data (black dots) and random samples (red dots) generated from the trained models: the proposed CRBM (above) and the conventional RBM (below).}
  \label{fig:comp}
\end{figure}

\subsection{Evaluation using artificial data}

In order to confirm the effectiveness of the proposed CRBM, we first conducted a simple experiment using one-dimensional complex-valued artificial data (the number of training data $N=2000$).
The artificially created data is illustrated in Fig.~\ref{fig:comp} as black dots, which has correlations between the real and imaginary parts.
In this experiment, we compared the CRBM to a conventional RBM that has two real-valued visible units; one is for the real part, another is for the imaginary part.
We trained both models with two hidden units using the steepest gradient ascent with a learning rate of $0.01$, a momentum of $0.1$, a batch size of $20$, and a number of epochs as $200$.
After the training, we randomly generated samples from the models; the samples from the CRBM are shown as red dots on the top of Fig.~\ref{fig:comp} and the RBM are shown as red dots on the bottom of Fig.~\ref{fig:comp}.
As shown in Figure~\ref{fig:comp}, we can see that the proposed CRBM could represent the distribution of the complex-valued artificial data more accurately than the RBM.
This is because the CRBM captures the relationships between the real and imaginary parts while the conventional RBM does not. 

\subsection{Evaluation using speech data}

Secondly, we conducted speech encoding experiments using speech signals of $50$ sentences (approx. $4.2$ min) for training and another $53$ for tests pronounced by a female announcer (``FTK'') from the set ``A'' of the ATR speech corpora.
The speech signals were downsampled from the original 20kHz to 16kHz, and processed into $129$-dimensional complex spectra using the short-time Fourier transform (STFT) with a window length of $256$ and a hop size of $64$, followed by the CPCA to obtain complex-valued features.
The total number of the training data was $64,438$.
In order to decide how much to reduce the number of dimensions by the CPCA, we examined the perceptual evaluation of speech quality (PESQ) of the recovered signals using the inverse short-time Fourier transform (ISTFT) and the overlap-add method from the CPCA features by changing the number of dimensions $P$ to $20$, $40$, $60$, $80$, and $100$, as shown in Table~\ref{tab:pca_pesq}.
Table~\ref{tab:pca_pesq} shows that the PESQ with $P=40$ is similar to those with higher $P$; therefore, we used $P=40$ in the rest of our experiments in terms of sufficient quality and dimensional reduction.

\begin{table}[t]
  \caption{PESQ of the reconstructed speech from CPCA-features.}
  \label{tab:pca_pesq}
  \centering
  \begin{tabular}{lrrrrrr}
    \toprule
    $P$ & \textbf{20} & \textbf{40} & \textbf{60} & \textbf{80} & \textbf{100} \\
    \midrule
    \textbf{PESQ}         & $3.71$     & $4.46$       & $4.49$     & $4.50$      & $4.50$ \\
    \bottomrule
  \end{tabular}
\end{table}

\begin{table}[t]
  \caption{PESQ of the CRBM and RBM methods when changing the number of hidden units (the leftmost column). The methods with the notation ``+T'' use trajectory estimation; otherwise, they use frame-wise estimation. ``-GL'' denotes the use of the Griffin-Lim algorithm.}
  \label{tab:res_h}
  \centering
  \begin{tabular}{lrrrrr}
    \toprule
    $H$ & \textbf{CRBM+T} & \textbf{CRBM} & \textbf{RBM+T} & \textbf{RBM} & \textbf{RBM-GL}\\
    \midrule
    \textbf{1k} & $\bs{2.41}$ & $2.34$ & $2.39$ & $2.30$ & $2.33$ \\
    \textbf{2k} & $\bs{2.72}$ & $2.60$ & $2.62$ & $2.54$ & $2.46$ \\
    \textbf{4k} & $\bs{2.81}$ & $2.70$ & $2.66$ & $2.54$ & $2.39$ \\
    \bottomrule
  \end{tabular}
\end{table}

\begin{figure}[t]
  \centering
  \includegraphics[width=0.90\columnwidth]{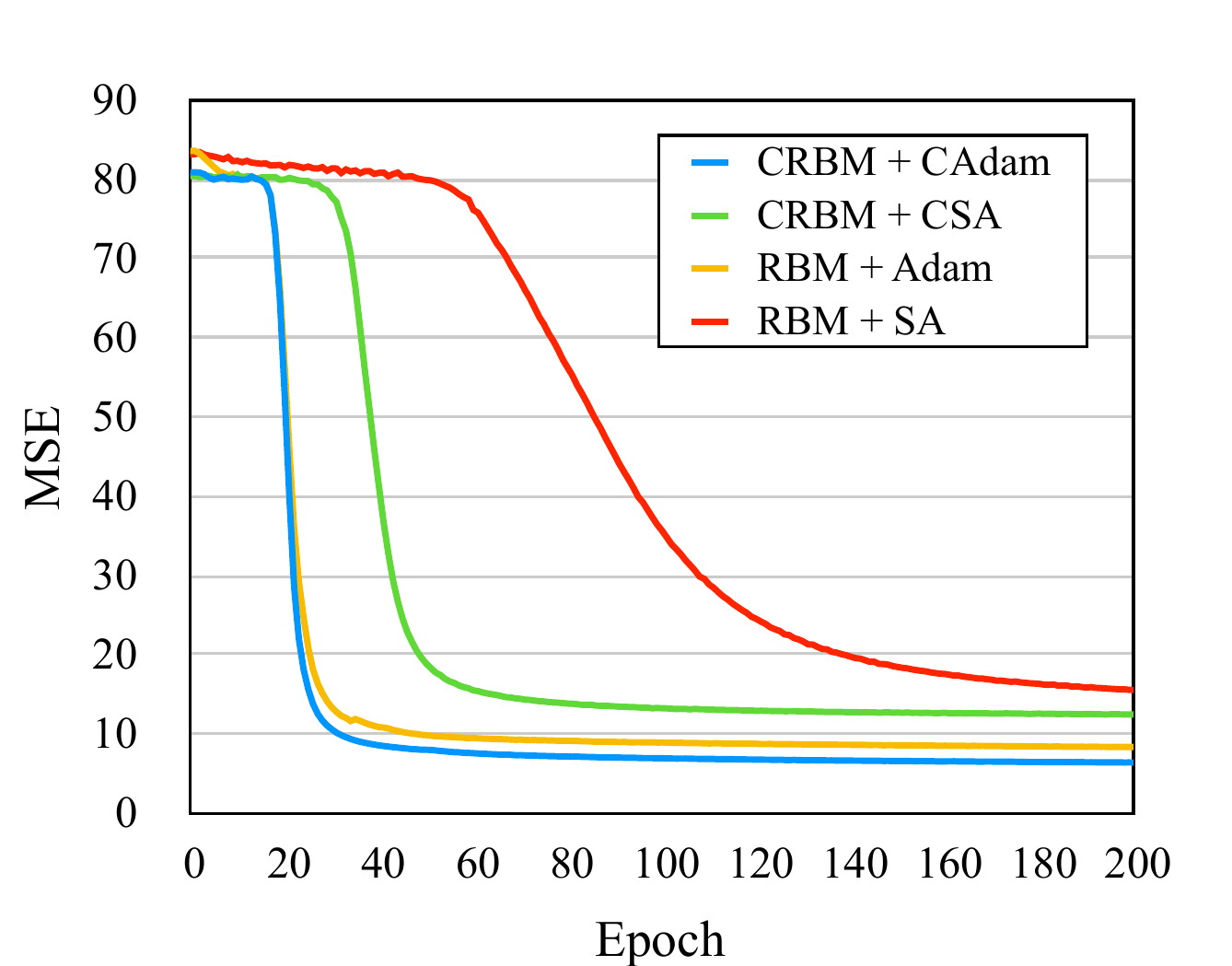}
  \caption{MSE curve during the training of the CRBMs and the RBMs.}
  \label{fig:mse}
\end{figure}

\begin{table}[t]
  \caption{Speech coding performance (PESQ) of each method. The notations are the same as in Table~\ref{tab:res_h}.}
  \label{tab:res}
  \centering
  \begin{tabular}{lr}
    \toprule
    \textbf{Methods} & \textbf{PESQ} \\
    \midrule
    \textbf{CRBM+T} & $\bs{2.81}$ \\
    \textbf{CRBM} & $2.70$ \\
    \textbf{RBM+T} & $2.66$ \\
    \textbf{RBM} & $2.54$ \\
    \textbf{RBM-GL} & $2.46$ \\
    \midrule
    \textbf{MCEP} & $2.68$ \\
    \textbf{CEP} & $2.54$ \\
    \textbf{WORLD} & $\bs{2.86}$ \\
    \bottomrule
  \end{tabular}
\end{table}

\begin{table}[t]
  \caption{PSNR [dB] of reconstructed spectra from the CRBM and RBM with respect to magnitude spectrum (MS) and phase difference (PD).}
  \label{tab:psnr}
  \centering
  \begin{tabular}{lrr}
    \toprule
    & \textbf{MS} & \textbf{PD} \\
    \midrule
    \textbf{CRBM} & $\bs{39.8}$ & $\bs{7.04}$\\
    \textbf{RBM} & $38.8$ & $6.72$\\
    \bottomrule
  \end{tabular}
\end{table}

\subsubsection{Methods compared}

We compared the proposed method (``CRBM+T'') to our previous model \cite{nakashika2017complex} (``CRBM''), the RBM that feeds concatenated real-valued vectors of real and imaginary parts of the CPCA features (``RBM''), and its trajectory version that feeds static and dynamic features (``RBM+T'').
Another RBM-based method (``RBM+GL'') we compared was trained using $40$-dimensional real-valued features obtained by amplitude spectra followed by PCA as visible units and recovered speech signals using the Griffin-Lim algorithm \cite{Griffin:1984bl}.
These models were evaluated by changing the number of hidden units $H$ to $1,000$, $2,000$, and $4,000$.
The CRBMs were trained using the stochastic gradient method of $100$-size mini-batches and $200$ epochs with a learning rate $\alpha = 0.01$ for the CSA and $\alpha = 0.001$, $\beta_1 = 0.9$, $\beta_2 = 0.999$ for the complex Adam (CAdam).
We set the same parameters for the RBMs except for using the real-valued steepest ascent (SA) and Adam.
For the gradient method to estimate the sequence in Eq.~\eqref{eq:update_z}, we used the CSA (SA for the RBM method) of $100$ epochs with a learning rate of $0.01$.
We also compared the proposed method to the traditional speech coding of cepstral (``CEP'') and mel-cepstral (``MCEP'') analysis.
The cepstral coefficients were $40$ and recovered speech using the log magnitude approximation (LMA) filter \cite{imai1978speech}.
From $20$-dimensional mel-cepstral coefficients, we restored speech using the mel-log spectrum approximate (MLSA) filter \cite{imai1983mel}.
Finally, we compared it to the WORLD \cite{morise2016world} as a high-quality speech analysis-by-synthesis system.

\begin{figure*}[t]
  \centering
  \includegraphics[width=2.00\columnwidth]{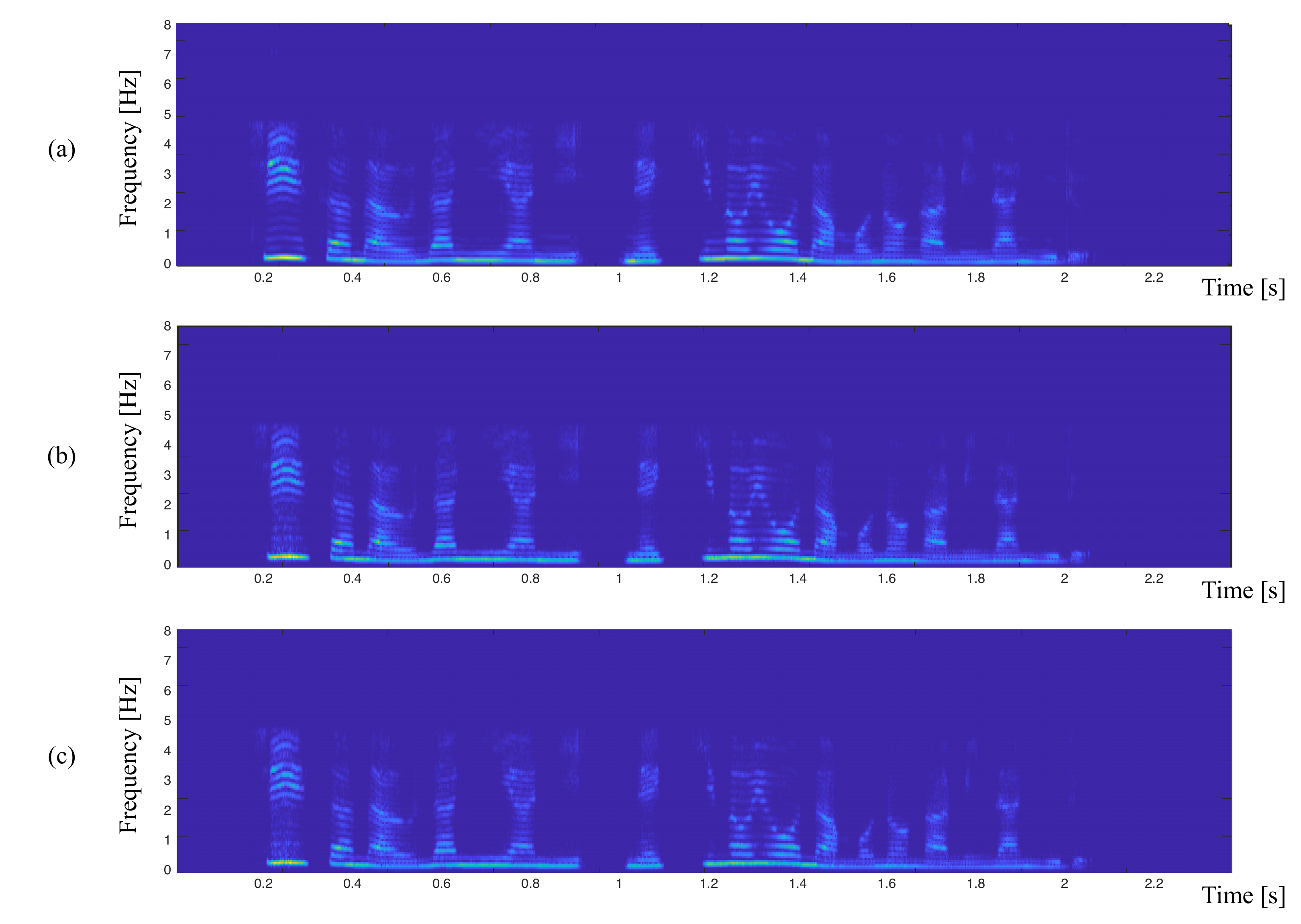}
  \caption{Magnitude spectra of the original speech (a), reconstructed from the CRBM (b), and reconstructed from the RBM (c).}
  \label{fig:spec}
\end{figure*}

\subsubsection{Objective evaluation}

Fig.~\ref{fig:mse} shows the mean-squared error (MSE) calculated during the training, comparing the CRBM with CAdam (``CRBM+CAdam'') to the counterparts.
As shown in Fig.~\ref{fig:mse}, the CRBMs converged more quickly than the counterparts of RBMs, and the CAdam algorithm was considerably effective for the CSA.
Table~\ref{tab:res_h} illustrates the performance of the CRBM and RBM methods in the test set, showing that the proposed method outperformed the rest regardless of the number of hidden units.
While the performance of the CRBMs with $H=1,000$ was comparable to that of the RBMs with $H=1,000$, the CRBMs with more hidden units highly improved the performance compared to the RBMs.
This is because each hidden unit in the CRBMs represents complex-valued patterns more independently than that in RBMs; i.e., CRBMs have a higher ability of representing complex-valued data than RBMs.

Table~\ref{tab:res} summarizes the performance of each method under their best conditions.
All methods based on CRBM and RBM were trained using the CAdam and Adam algorithms.
Interestingly, the CRBM with frame-wise modeling (``CRBM'') outperformed the RBM with trajectory modeling (``RBM+T'') because the CRBM implicitly represents the phase information of complex-valued data, and the frame-wise features from the CRBM recovered speech sufficiently.
Furthermore, the proposed trajectory modeling (``CRBM+T'') improved accuracy by extracting the correlations of complex-valued features between adjacent frames and performed the best out of all the training-based methods.
The performance of the proposed method is even comparable to that of the WORLD, which is one of the most high-quality synthesis methods.
Unlike traditional speech coding (mel-cepstrum, cepstum, and WORLD), the CRBMs directly encode arbitrary complex-valued features into binary-valued features, which indicates that the CRBMs have a high compression ability and high versatility to complex-valued data and can be applied to speech and to other signals.

Table~\ref{tab:psnr} compares the peak signal-to-noise ratio (PSNR) of the CRBM and RBM in order to analyze which magnitude or phase in the reconstruction of the CRBM is actually effective.
According to Table~\ref{tab:psnr}, the CRBM got 2.58\% relative improvement points to the RBM in terms of magnitude and 4.76\% relative improvement points to the RBM in terms of phase.
This demonstrates that the CRBM can effectively represent complex-valued data in particular with respect to phase.
Therefore, although the magnitude spectra of the CRBM reconstruction are very similar to those of the RBM (as shown in Fig.~\ref{fig:spec}), the generated speech from the CRBM outperformed that from the RBM in terms of the PESQ, as shown in Table~\ref{tab:res}.

\begin{figure}[t]
  \centering
  \includegraphics[width=1.00\columnwidth]{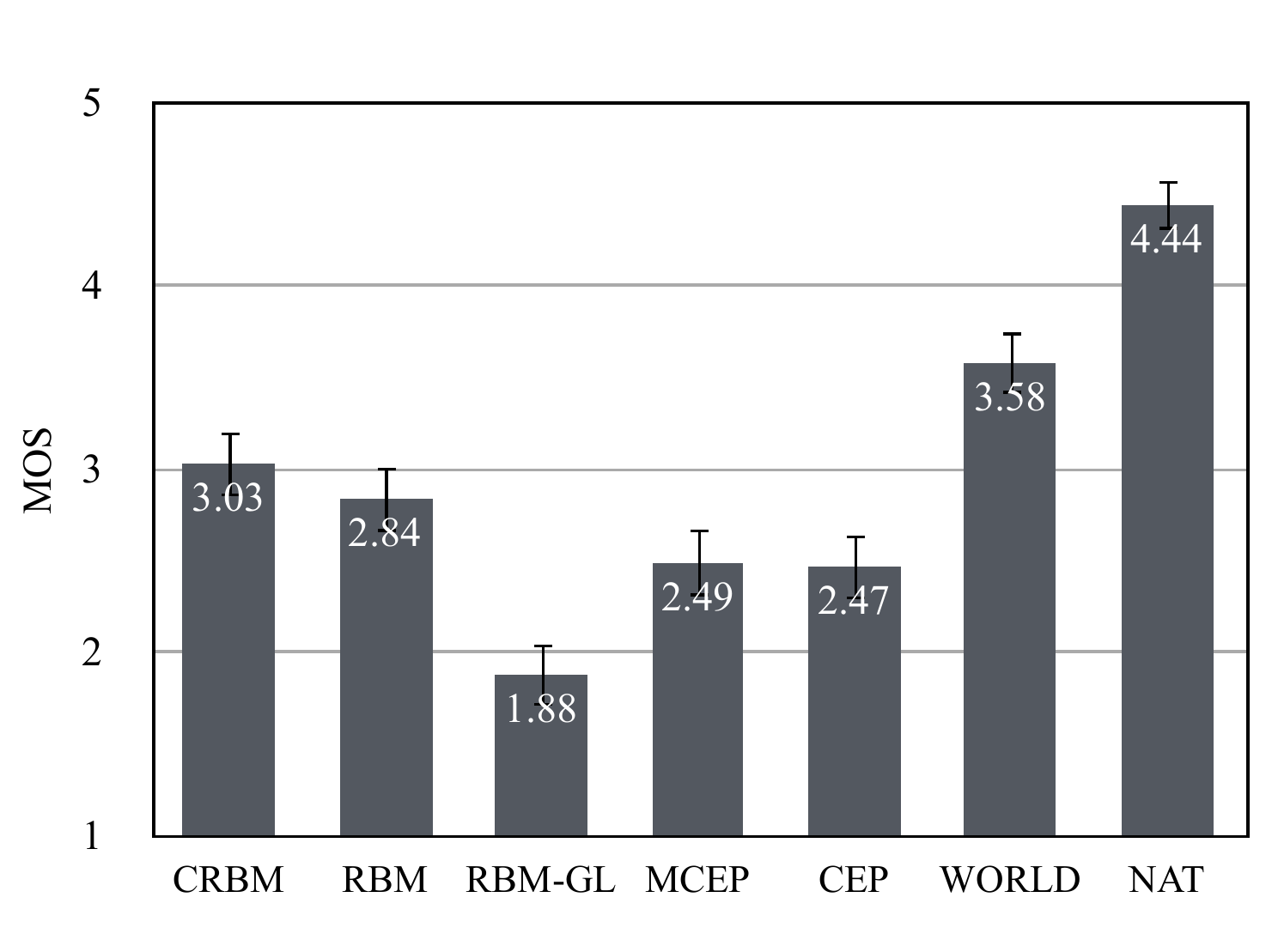}
  \caption{Mean opinion score (MOS) for each method. The numbers in bars indicate the average of the MOS and the error bars indicate the confidential intervals.}
  \label{fig:mos}
\end{figure}

\subsubsection{Subjective evaluation}

Finally, we conducted subjective experiments based on the mean opinion score (MOS) of $95$ participants gathered through crowdsourcing.
Each participant was asked to rank the synthesized and natural speech (NAT) on a 5-point scale (1: poor, 2: fair, 3: good, 4: very good, and 5: excellent) in terms of speech quality (naturalness).
Because the MCEP, the CEP, and the WORLD are based on frame-wise synthesis, we used frame-wise estimation for the CRBM and the RBM rather than trajectory estimation in these experiments.
Fig.~\ref{fig:mos} shows the results of the subjective evaluation.
As shown in Fig.~\ref{fig:mos}, the CRBM performed the best out of all the methods except the WORLD.
We also conducted pairwise t-tests for each combination and observed significant differences with a 95\% confidence for all pairs except for the difference between the CEP and MCEP.

\section{Conclusion}
\label{sec:conc}

We proposed a ``complex-valued RBM'' (CRBM), a novel probabilistic model that extends an RBM in order to feed complex-valued data.
This paper also includes its improved learning methods in modeling speech: the dimensionality-reduction of complex-valued data using CPCA, the CAdam learning algorithm to estimate complex-valued parameters more effectively, and the trajectory modeling and the generation method of complex-valued data based on MLPG.
Experimental results showed the effectiveness of the proposed method with objective and subjective criteria compared to the other speech coding methods except the WORLD.
Although the CRBM fell just one step short of the WORLD in terms of quality as a specialized coding method for speech, the CRBM can be also used for coding other signals such as music, images, array signals, etc.
We will further investigate the high ability of the CRBM in such applications in the future.

We presented the CRBM in this paper as a very basic model and believe that the model can be easily extended.
For example, we could define extensions by stacking two or more hidden layers like the deep Boltzmann machine \cite{salakhutdinov2009deep} by adding connections from the previous to the current hidden/visible units like the recurrent temporal Boltzmann machine \cite{sutskever2009recurrent}, or by changing the energy function to form the conditional probability of hidden units such as Gaussian distribution or complex normal distribution, etc.
The deep extension can be also used as a pre-training method for complex neural networks \cite{nemoto1992complex}.
Future work includes such extensions.


%



\section*{Acknowledgment}
This work was partially supported by JST ACT-I, by MEXT KAKENHI Grant Numbers (15H01686, 16K16096, 16H06302,17H04687), and by the Telecommunications Advancement Foundation Grant.


\ifCLASSOPTIONcaptionsoff
  \newpage
\fi



%



\bibliographystyle{IEEEtran}
\bibliography{crbm}

%

\begin{IEEEbiography}{Toru Nakashika}
received his B.E. and M.E. degrees in computer science from Kobe University in 2009 and 2011, respectively. In the summer in 2010, he was a student researcher at IBM Research, Tokyo Research Laboratory. From September 2011 to August 2012, he was a visiting researcher in the image group at INSA Lyon in France. In that year, he continued his research as a doctoral student at Kobe University and received his Dr.Eng. degree in computer science in 2014. He was an Assistant Professor at Kobe University until April 2015. He is currently an Assistant Professor at the University of Electro-Communications in Chofu, Japan. He received the IEICE ISS Young Researcher's Award in Speech Field in 2013 and the IPSJ Ongaku Symposium Excellent Paper Award in 2016. He is a member of IEEE, IEICE, and ASJ.
\end{IEEEbiography}

\begin{IEEEbiography}{Shinji Takaki}
 received a B.E. degree in computer science and received an M.E. and a Ph.D. degree in scientific and engineering simulation from the Nagoya Institute of Technology, Nagoya, Japan in 2009, 2011, and 2014, respectively. From September 2013 to January 2014, he was a visiting researcher at the University of Edinburgh University. Since April 2014, he has been a project researcher at the National Institute of Informatics in Japan. His research interests include statistical machine learning and speech synthesis. He is a member of the Acoustical Society of Japan and the Information Processing Society of Japan.

\end{IEEEbiography}

\begin{IEEEbiography}{Junichi Yamagishi}
(SM'13) is an associate professor of the National Institute of Informatics in Japan. He is also a senior research fellow at the Centre for Speech Technology Research (CSTR) at the University of Edinburgh, UK. He was awarded a Ph.D.\ by the Tokyo Institute of Technology in 2006 for a thesis that pioneered speaker-adaptive speech synthesis and was awarded the Tejima Prize for the best Ph.D.\ thesis of the Tokyo Institute of Technology in 2007. Since 2006, he has authored and co-authored over 100 refereed papers in international journals and conferences. He was awarded the Itakura Prize from the Acoustic Society of Japan, the Kiyasu Special Industrial Achievement Award from the Information Processing Society of Japan, the Young Scientists’ Prize from the Minister of Education, Science, and Technology, and the JSPS prize in 2010, 2013, 2014, and 2016, respectively. 

He  was one of organizers for special sessions on ``Spoofing and Countermeasures for Automatic Speaker  Verification'' at  Interspeech  2013,  ``ASVspoof evaluation''  at  Interspeech  2015, ``Voice conversion challenge 2016'' at Interspeech 2016, and ``2nd ASVspoof evaluation'' at Interspeech 2017.  He  has been  a member of  the  Speech  \& Language  Technical  Committee  (SLTC). He served as an Associate  Editor  of  the IEEE/ACM Transactions on Audio, Speech, and Language Processing and as a Lead Guest Editor for the IEEE Journal of Selected Topics in Signal Processing (JSTSP) special issue on Spoofing and Countermeasures for Automatic Speaker Verification.
\end{IEEEbiography}




\end{document}